\documentclass[sigconf]{acmart}

\usepackage{threeparttable}
\usepackage{subcaption}
\usepackage{todonotes}
\usepackage{comment}

\AtBeginDocument{%
  \providecommand\BibTeX{{%
    \normalfont B\kern-0.5em{\scshape i\kern-0.25em b}\kern-0.8em\TeX}}}

\setcopyright{acmcopyright}
\copyrightyear{2021}
\acmYear{2021}
\acmDOI{10.1145/1122445.1122456}

\acmConference[CHI '21]{CHI '21: ACM CHI Conference}{May 08--13, 2021}{Yokohama, Japan}
\acmBooktitle{CHI '21: ACM CHI Conference,
  June 08--13, 2021, Yokohama, Japan}
\acmPrice{15.00}
\acmISBN{978-1-4503-XXXX-X/18/06}

\begin{document}

\title{VINS: Visual Search for Mobile User Interface Design}

\author{Sara Bunian}
\affiliation{%
  \institution{Northeastern University}
  \city{Boston}
  \state{MA}
  \country{USA}
}
\email{banian.s@northeastern.edu}

\author{Kai Li}
\affiliation{%
  \institution{Northeastern University}
  \city{Boston}
  \state{MA}
  \country{USA}
}
\email{kaili@ece.neu.edu}

\author{Chaima Jemmali}
\affiliation{%
  \institution{Northeastern University}
  \city{Boston}
  \state{MA}
  \country{USA}
}
\email{jemmali.c@northeastern.edu}

\author{Casper Harteveld}
\affiliation{%
  \institution{Northeastern University}
  \city{Boston}
  \state{MA}
  \country{USA}
}
\email{c.harteveld@northeastern.edu}

\author{Yun Fu}
\affiliation{%
  \institution{Northeastern University}
  \city{Boston}
  \state{MA}
  \country{USA}
}
\email{yunfu@ece.neu.edu}

\author{Magy Seif El-Nasr}
\affiliation{%
  \institution{University of California at Santa Cruz}
  \city{Santa Clara}
  \state{California}
  \country{USA}
}
\email{mseifeln@ucsc.edu}

\renewcommand{\shortauthors}{Trovato and Tobin, et al.}

\begin{abstract}
Searching for relative mobile user interface (UI) design examples can aid interface designers in gaining inspiration and comparing design alternatives. However, finding such design examples is challenging, especially as current search systems rely on only text-based queries and do not consider the UI structure and content into account. This paper introduces VINS, a visual search framework, that takes as input a UI image (wireframe, high-fidelity) and retrieves visually similar design examples. We first survey interface designers to better understand their example finding process. We then develop a large-scale UI dataset that provides an accurate specification of the interface's view hierarchy (i.e., all the UI components and their specific location). By utilizing this dataset, we propose an object-detection based image retrieval framework that models the UI context and hierarchical structure. The framework achieves a mean Average Precision of 76.39\% for the UI detection and high performance in querying similar UI designs.
\end{abstract}



\begin{teaserfigure}
  \includegraphics[width=\textwidth]{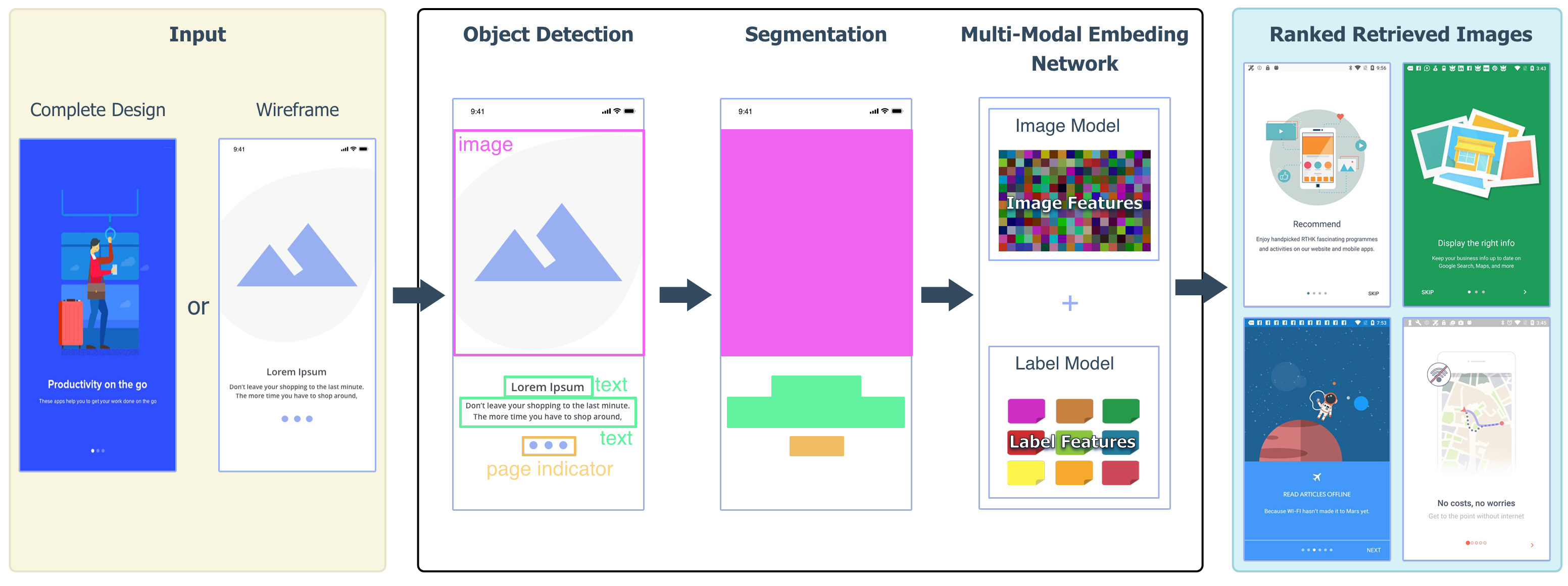}
  \caption{Overview of \textit{VINS}, our proposed image retrieval process for the visual search for mobile interface design. First, it takes as \textit{input} UI layout screens, either a complete design or an abstract wireframe. Then, it employs an \textit{object detection} model to detect the presence and location of the different UI components defining the input query and produces a \textit{segmented layout} accordingly. This segmented layout is passed to a \textit{multi-modal embedding network} that learns a joint feature representation of both visual and label features. This representation is used to retrieve a \textit{ranked list} of similar designs.}
  \Description{The process of VINS starts with taking an app's layout as its input query, which can be of different representations, including abstract wireframes and high-fidelity layouts. It then goes through a series of operations including detection, segmentation, and retrieval through a multi-modal embedding network. Based on the outcome of this process it retrieves finds the most similar matches to the input query from our design inventory.}
  \label{fig:overview}
\end{teaserfigure}

\maketitle

\section{Introduction}
Today's digital, fast-paced world has stimulated a rapidly growing mobile application (also known as apps) industry. Within the mobile app design process, the User Interface (UI) is an important visual communication factor that can play a significant role in the app's success. The UI depicts the organization and visual structure of the different components comprising the app's layout. \textit{Design examples} are important in the UI design process~\cite{swearngin2018rewire}, and interface designers usually search and create design example repositories to stimulate inspiration, generate new ideas, and investigate feasible options to make design decisions~\cite{bonnardel1999creativity,herring2009getting,herring2009idea,buxton2010sketching}.

The web is presently a large repository containing a collection of design examples. There are various specific online design sharing websites that provide UI design inspiration, such as \textit{uplabs}\footnote{https://www.uplabs.com/} and \textit{dribbble}\footnote{https://dribbble.com/}. However, the current searching mechanism within these websites is similar to the web at large, limited to only text-based queries, which makes finding relative examples a challenging task~\cite{herring2009getting,miller2014searching,sharmin2009understanding}. While designers can easily search using keywords (e.g., onboarding screens), general categories (e.g., Food Apps), and color codes (e.g., white interface), the resulting design examples might not be relevant to the original design requirements in terms of visual layout structure and UI content. This makes the searching process tedious and slow. To address this issue, more effective tools are needed to support finding and retrieving design examples that will benefit designers in practice. 

Typically, designers express their ideas and UI design concepts in the form of images that describe the UI's visual layout, hierarchical structure, and content. There has recently been a growing interest to study mobile UI retrieval based on an input image~\cite{huang2019swire,liu2018learning}. However, the proposed methods so far present limitations in terms of performance and generalizability. For example, \textit{Swire}~\cite{huang2019swire} does not specifically consider the UI content in the retrieval process. This affects the performance of the system by retrieving images that are not relevant to the query or are missing design components. The approach presented by Liu et al.~\cite{liu2018learning} depends solely on a predefined UI content hierarchy. This limits the generalizability of the approach and prevents it from working on any new unseen images. Therefore, there is a need for a fine-grained visual search that, given any input query, can infer the UI's content hierarchy and provide designers with examples that fit their query.

In this paper, we take a step towards investigating how to effectively address the process of image retrieval in the domain of mobile UI design. Our approach focuses on two main aspects for the retrieval process: First, it is essential to have an understanding of the collection of UI components and their spatial location in a UI design image. This provides a close approximation to the UI visual hierarchy which allows measuring the relevance across different images at a more fine-grained level. Second, the system should be flexible to allow designers to find similar design examples for their UI query in different design stages. This provides flexibility in either using an abstract wireframe (i.e., low/medium-fidelity image) or a UI screenshot (i.e., high-fidelity design image) as in input query. 

In order to develop VINS, our proposed visual search framework, we first report findings of semi-structured 
interviews with UI designers explaining the current process and informing the design requirements of UI searching tools. We then explore the issue of UI image retrieval in detail from both the data and the model side. Figure~\ref{fig:overview} shows VINS, our proposed visual search framework, which takes an app's layout as its input query and finds the most similar matches from our design inventory. To account for various stages in the design process, the input query can be of different representations, including abstract wireframes and high-fidelity layouts. To support this feature, we constructed a large-scale annotated dataset containing UI design screens across these two design stages that we refer to as VINS dataset. 

Given the variety of the UI design screens and the complexity of their visual hierarchy, we develop VINS around deep learning models, which have demonstrated their effectiveness in solving various tasks in different contexts ~\cite{ dargan2019survey}. Specifically, VINS consists of two building blocks: detection and retrieval. First, we utilize an object detection mechanism to detect the presence and location of the different UI components that represent a tentative layout of the input query. We then train an attention-based neural network to learn a joint feature representation that can define both the layout structure and its content in order to retrieve similar UI designs.

The paper provides the following contributions:

\begin{enumerate}

\item VINS dataset: A large mobile UI dataset consisting of UI screens across different design stages (i.e., abstract wireframes and high-fidelity designs) that can be utilized in developing different data-driven design applications. Through a human-powered process, we annotate the dataset to provide an accurate specification of the UI in terms of its view hierarchy (i.e., all the various UI components and their specific location). 

\item VINS: A deep learning framework that models the context and the hierarchical structure of UI screens to develop a UI image retrieval system. The framework can achieve high performance in querying similar UI designs. 

\end{enumerate}
\section{Related Work}
\subsection{Visual Search}
Current design search systems encompass a variety of domains including web design~\cite{ritchie2011d,kumar2013webzeitgeist}, 3D modeling~\cite{funkhouser2003search}, interior design~\cite{bell2015learning}, fashion~\cite{mcauley2015image}, and programming~\cite{brandt2010example}. However, the current search mechanism in these and other systems is mostly based on text queries, such as keywords. Existing research has shown that keywords often fail to articulate the abstract design ideas and thus makes finding relevant design examples a challenging task for designers~\cite{herring2009getting,miller2014searching,sharmin2009understanding}.

Several research studies have examined how designers essentially search for design examples, and how these examples are utilized in supporting their creative process~\cite{herring2009getting,herring2009idea}. To support designers in this example finding task, a plethora of HCI research has investigated alternative ways to better explore and retrieve design examples. One approach is to use advanced keywords such as stylistic features (e.g., color, style term)~\cite{ritchie2011d,lee2010designing}, or image metadata (e.g., themes, media, date, location, shapes). Because formulation of search queries using images is easier to learn and faster to specify than keywords~\cite{yeh2009sikuli}, other researchers have explored alternative visual searching mechanisms using image queries such as sketches~\cite{hashimoto2005retrieving} and UI screenshots~\cite{yeh2009sikuli}. 

There has been a recent interest in advancing the body of visual search by integrating deep learning frameworks for better results. To this end, Deka et al.~\cite{deka2017rico} presented preliminary results of an autoencoder model that learns similarities between UI layouts based on image and text content only. Liu et al.~\cite{liu2018learning} extended this approach by training an autoencoder on semantically annotated layouts, containing different UI components, text button concepts, and icon classes, to learn UI similarities for design search. We also use an autoencoder for our retrieval task, however, our model is different because we develop an attention-aware autoencoder that learns a joint embedding of structure and content.

The most closely related approach to our work in terms of visual search is the aforementioned \textit{Swire} system~\cite{huang2019swire}, which uses a deep neural network model to retrieve relevant UI examples from input sketches. Specifically, \textit{Swire} trains two convolutional sub-networks over matching pairs of screenshots and their corresponding sketches. \textit{Swire} achieved 60\% relevancy of retrieved examples and demonstrated its applicability in a number of different tasks. However, it has the following key limitations: (1) focuses only on the high-level layout information without inferring the UI's structure and content, which sometimes retrieves images that are irrelevant or missing UI components; and (2) requires a pairwise collection of sketches and screenshots which makes it difficult to generalize across unseen sketches of UI layouts. We seek to address these limitations and advance the body of work on visual search with our approach (see Figure~\ref{fig:overview}).

\subsection{Mobile UI Datasets and Detection}
Early work on data-driven design explored how design examples can aid in various design tasks, including providing design assistance~\cite{swearngin2018rewire} and automated content and layout re-organization~\cite{kumar2011bricolage, o2014learning, o2015designscape}. Motivated by the utility of design examples in supporting designers and enabling data-driven application design, several large-scale mobile UI datasets have been created. For example, \textit{ERICA}~\cite{deka2016erica} provides a collection of user interaction data for mobile UIs captured while using the app. 

\begin{figure}
  \centering
  \includegraphics[width=\columnwidth]{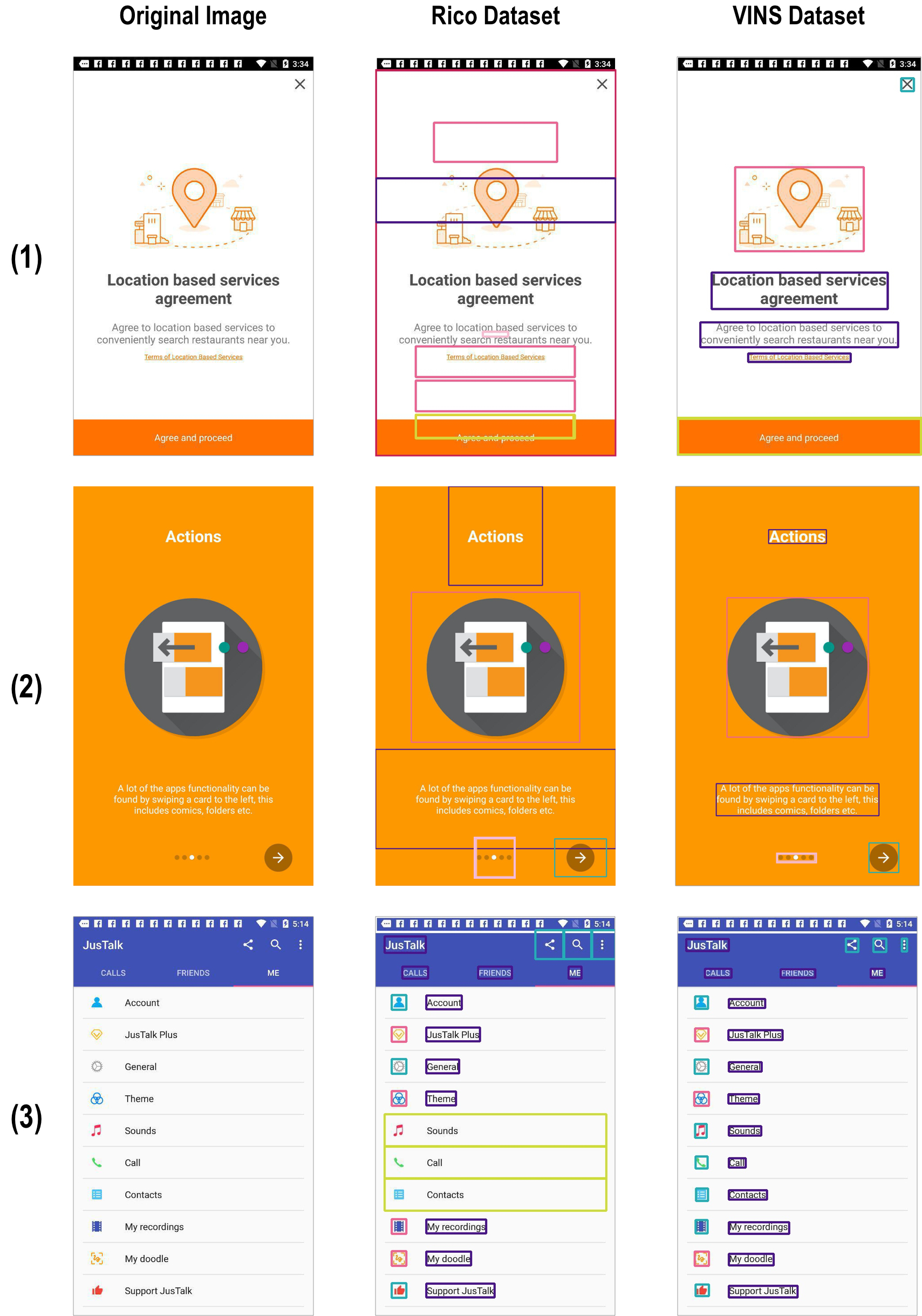}
  \caption{Comparison of bounding boxes annotations between \textit{Rico} and VINS dataset.}
  \Description{Comparing the bounding boxes between the predefined view hierarchies of the RICO dataset and the annotations of the VINS Dataset. We display 3 columns: the first one includes the original UI image; the rest 2 columns draw the bounding boxes of each dataset on the original image to demonstrate its accuracy. The figure includes 3 UI examples and we identify three limitations of Rico’s dataset which we were able to resolve with our annotations including (1) hierarchies may be broken and are not directly mapped to the UI, (2) additional space inside the bounding box boundaries does not reflect the exact dimensions and position of the object, and (3) inconsistencies with the class label of similar objects.}
  \label{fig:BBox_Comparison}
\end{figure}

More recently, \textit{Rico}~\cite{deka2017rico, liu2018learning}, a large-scale data of mined Android apps, has been released. It consists of 72K UI examples from 9,722 Android apps. Each example is associated with a screenshot of the UI design, the corresponding view hierarchy, and the user interaction information. The predefined view hierarchies expose the UI's structural and functional properties, which provides a means of inferring the UI content hierarchy that has the potential of supporting various data-driven design applications. While these view hierarchies may often provide an accurate representation of the UI structure, there are several instances where they do not. As shown in Figure ~\ref{fig:BBox_Comparison}, (1) hierarchies may be broken and are not directly mapped to the UI, (2) additional space inside the bounding box boundaries does not reflect the exact dimensions and position of the object, and (3) inconsistencies with the class label of similar objects. Hence, it is important to have a dataset that can provide a highly accurate content hierarchy. Such a dataset enables to train computer vision systems, e.g. object detector, to analyze relevant patterns and recognize objects.

Other researches have employed traditional image processing techniques (e.g., Optical Character Recognition (OCR)~\cite{nagy1999optical}, Canny Edge Detection~\cite{canny1986computational}) to infer the UI structure and content~\cite{nguyen2015reverse,moran2018machine,swearngin2018rewire,behrang2018guifetch}. However, this method is constrained in the identification of components that are cluttered with 
the background and involves the adoption of an image classification model to differentiate between components~\cite{moran2018machine}.

To effectively address the issue of inferring UI structure, an approach that can perform both the tasks of object localization and classification is needed. Object detection is one of the key problems in the computer vision community, which aims to provide a comprehensive understanding of the image by precisely determining the location and category of its objects through learning and extracting high-level deep visual features~\cite{zhao2019object}. Our method for inferring the UI's layout structure integrates an object detection mechanism to accurately detect the different UI components. Object detection has been used recently for describing the UI structure~\cite{yun2018detection}. This approach, however, was restricted to UI sketches and the dataset was not made public. To support this purpose, we are collecting VINS dataset, a new fully annotated dataset, which we will describe in Section~\ref{section:data}.
\section{Formative Interviews}
\label{section:interview}
We conducted a series of semi-structured interviews with UI designers to gain insight from a professional user's perspective into the current design process and help assess the applicability of VINS into the design workflow. A total of 24 designers were recruited from the freelancing website \textit{Upwork}\footnote{https://www.upwork.com/}. All designers were based in the USA and reported receiving a formal UI/UX training. The designer’s experience ranged from 1 to 5 years with an average of about 2 years. They were compensated \$10 USD and the interview took about 30 minutes. 

The interview script consisted of 23 questions related to (1) the current strategy of finding design examples; (2) the difficulties encountered throughout the process; (3) aspects of similarity between UI designs; and (4) the applicability of VINS. To analyze the survey questions, we performed a 3-stage thematic coding for analyzing results. We began by literally coding keywords of designers, then grouping the responses through discussion according to the themes that emerged, and finally reporting how many designers used the various themes. Two researchers engaged in this consensual qualitative process~\cite{hill2012consensual}, a third researcher confirmed the findings.   

In line with the findings reported by Herring et al.~\cite{herring2009getting,herring2009idea}, most designers (18 out of 24) agreed that design examples are often used for almost every project they work on. For example, as D8 explained, “I find them very helpful and tend to use them as much as possible. Every project of mine contains at least 60\% of design examples”. Similarly, D6 stated “Everytime [sic] when I have to design. It’s a compulsory thing for me”. Their role in the design process is therefore beneficial to the designers. Specifically, when asked if exposure to examples is a good means of inspiration, the answers were mostly favorable (“Definitely yes”: 15 designers, “Probably yes”: 5 designers, “Might or might not”: 3 designers, "Definitely not": 1 designer). 

Given the vast design space, designers further reported that they collect different design aspects from examples, such as various layouts (22 designers), font styles (14 designers), different palettes (11 designers), and content hierarchy (4 designers). As D6 mentioned, “Mostly the layouts, fonts and how to organize data on the screen”. Similarly, D11 mentioned “Mostly it’s the layout. Other than that color schemes, font styles also”. In addition to the common themes, some designers mentioned that examples allow them the potential to: (1) explore emerging design trends: “It gives us an idea of what type of design trends now a day and explores many designs widgets” (D12); (2) discover new ideas: “Examples give you different ideas to create and manage your content” (D2); (3) increase creativity: “By looking at the designs creativity can be increased which can help you creating new designs of your own” (D7); and (4) identify visual attractiveness: “It helps you find what looks attractive and what's not” (D17).

By analyzing the themes of how these examples are collected, we found that designers often utilize different strategies. Most commonly, 21 designers reported that they search for these examples using keywords either by browsing the web (i.e., google) or navigating through various design sharing websites (i.e., Behance, Pinterest). As D1 mentioned “Yeah, I would look out on Google, Behance to search the best design examples”. Similarly, D4 mentioned “I search on Google and design websites such as Free pic, Pinterest with keywords”. Other designers would survey the market to view what has already been done by their competitors or utilize their old work as inspirational examples. As D6 stated “Going through the online UI kits for inspiration and sometimes I use my old designs”. 
 
Although the process of finding design examples is beneficial, it is also tedious and time-consuming. While 8 designers mentioned that it depends on what designs they are searching for, most designers stated that it usually takes a long time.  Specifically, 9 designers mentioned it usually takes hours to find good examples, while 3 designers stated that sometimes it may take more than a day. As D11 stated “Usually, it takes about half a day (4-5 hours) to find all the relevant design examples”. Also, D20 mentioned “It depends on what type of design you find but it takes 1 day”. In addition, designers identified other key obstacles they face within the process. As part of the current searching mechanism, 21 designers indicated that they use keywords to find design examples. Although keywords are simple, they have certain limitations. For example, keywords are limited in their ability in describing the design specifications: “searching results doesn’t meet our specific design requirements” (D2). More specifically, it is difficult to describe the specific layout design with only text queries: “don't know what exact query I should write” (D17) and “difficulty to find the right description of what's in my mind” (D21). As a result, the resulting design examples from the search might not be relevant to the original design requirements specified in the query. As D24 said: “I personally find this impossible to do. I may or may not be able to find an inspiration with the layout I have in mind. In this case, I usually only search for design inspirations to pick out a color scheme that I'll implement to the design layout which I already have in mind”. 

Although the current searching process is currently limited by keywords, designers did not completely agree on the role of keywords to effectively retrieve similar layouts. In response to whether UI similarity can be measured by keywords, 10 designers reported that they can be considered an indication of similarity, while 14 designers pointed that other design aspects should be considered. As D11 stated “Keywords are not the only measure for design similarity. Functionality and structure help in it too”. Similarly, D12 also supported this opinion by stating “Not only depends on keywords but also may be its functionally and visuality are not the same”. This emphasizes the need to consider new design aspects, other than keywords, for developing effective tools that support retrieving similar design examples.

To have a better understanding of the definition of layout similarity from a professional perspective, we asked designers to elaborate on their definition of similarity in regard to three aspects: structure, functionality, and visual elements and to identify which of these aspects is more important in the retrieval process. Designers have different perspectives regarding these aspects and their importance. Specifically, 7 designers agreed that all three aspects are equally important and play a key role within the searching mechanism, as D5 stated “The structure is first and main thing in design and visual is another thing that attract us. But functionality is important, but it will be according to design requirements”. The other remaining designers favored particular aspects as D11 mentioned “I find those designs helpful that have a functional and structural similarity with what I need”. D21 also supported this similarity definition “I would categorize two mockups that have a similar structure and functionalities but different color schemes to be `similar UI layouts'”. 

Thus, our formative interviews confirm what we found in the literature about the limitations of the use of keywords. To overcome this limitation, we also identify the importance of considering the new design aspects of functionality, structure and visual elements in order to better support the example finding process. In contrast to previous work ~\cite{huang2019swire, liu2018learning}, in VINS we emphasize functionality and structure in our design search. 
\section{Mobile UI Dataset}
\label{section:data}
Our approach for retrieving similar UI designs is based on object detection. The goal of object detection is to detect all instances of objects from a given set of classes and localize their exact positions in the image. The location of an object is defined in terms of a bounding box, which is represented by the rectangular boundary coordinates that fully enclose the object. 

Typically, training a good detector requires having a large number of training images in which objects are annotated with high-quality bounding boxes~\cite{dalal2005histograms, everingham2010pascal, felzenszwalb2009object, girshick2014rich}. For the development of VINS, it is thus essential to have a large-scale, carefully annotated dataset of UI design screens. To the best of our knowledge, there is no large-scale public dataset available that serves this purpose. Therefore, we created VINS dataset\footnote{https://github.com/sbunian/VINS}, a new annotated dataset containing a representative collection of UI screens across two design stages: abstract wireframes and high-fidelity fully designed interfaces. All of these UIs are annotated with bounding boxes spanning different classes of UI components. We identified a total of 11 UI components with varying functionality: background images, sliding menus, pop-up windows, input fields, icons, images, texts, switches, checked views, text buttons, and page indicators. Based on our analysis and due to relatively small training instances, we combined radio buttons and checkboxes to represent the checked view class. 

\subsection{Data Collection}
VINS dataset has a total of 4,800 images of UI designs screens, including 257 images of abstract wireframes and 4,543 images of high-fidelity screens. We opted to include images of different design stages to ensure that the VINS can perform on a wider variety of design inputs.

\subsubsection{Wireframes}
\begin{figure}
  \centering
  \includegraphics[width=1.0\columnwidth]{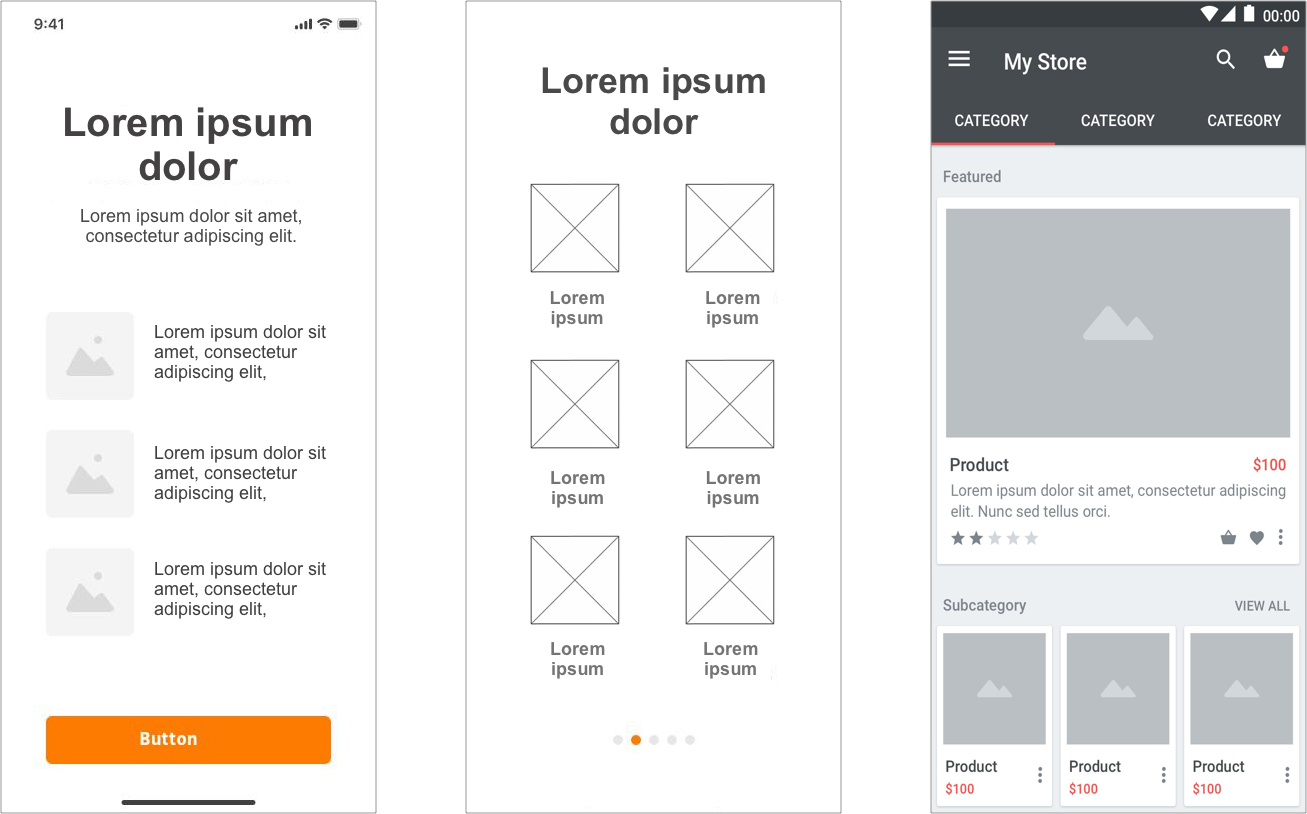}
  \caption{Wireframe templates with different prototyping styles.}
  \Description{This figure displays 3 examples of wireframe templates of different prototyping styles such as representing an image placeholder with either a mountain, or a square with a cross.}
  \label{fig:Wireframe}
\end{figure}

The wireframe-based dataset represents the initial stage of design and contains digital low/medium-fidelity images that describe the outline of the UI screen. From \textit{Uplabs}, we collected 257 abstract wireframe designs of different templates and layouts. We only collected a relatively small subset of images because of the simplicity of these wireframes, which represent the skeleton of the interface and are typically stripped from all styling and design elements that might affect the detection process. Wireframes can be generated using different tools, including whiteboard, paper-and-pencil, and a graphic design application. Because designers use different prototyping styles as shown in Figure ~\ref{fig:Wireframe}, such as representing an image placeholder with either a mountain, or a square with a cross, we included wireframe templates that represent different prototyping styles, thereby ensuring the generalizability of VINS.

\subsubsection{High-Fidelity UI Screens}
The high-fidelity-based dataset contains 4,543 images of carefully selected quality UI designs. To ensure that VINS generalizes across different platforms, we included images of both iPhone and Android interfaces from popular apps across different categories. For Android, we first started by manually selecting 2,000 high quality screens from the \textit{Rico} dataset~\cite{liu2018learning}. Due to \textit{Rico}'s large scale, it was very difficult to filter the duplicate, non-English, and outdated UIs. As a result, we additionally collected 740 UI images by navigating through different popular Google Play apps and taking screenshots of the UIs resulting in a total of 2,740 Android screens. To learn the iPhone design patterns, we downloaded and browsed a number of apps from different categories and took a screenshot of each screen, resulting in a total of 1,200 UI images. To ensure quality selection in both Android and iPhone platforms, we ensured that the selected UIs are from popular apps across different categories having an average rating more than 4 stars. To allow the system to identify new design trends, we collected 603 UI designs from the community-powered website \textit{Uplabs}, which offers quality digital UI inspirations from different designers. To find these images we used several keywords such as “Mobile UI Kit”, “Mobile Onboarding Screen”, “Mobile Login screen”, etc.

\subsection{Annotation Process}
Crowdsourcing~\cite{su2012crowdsourcing} for data annotation has recently attracted a lot of attention within the computer vision community due to the need for large-scale data and the lack of sufficiently labeled data. To this end, we utilized crowdsourcing where we recruited 6 students to thoroughly annotate VINS dataset. The students were recruited by the university’s internal slack channel and they were compensated with \$12 USD per hour. 

Inspired by the approach presented by Su et al. ~\cite{su2012crowdsourcing}, we follow a similar strategy to crowd-source bounding box annotations. The goal of this strategy is to ensure the bounding box's high quality and complete coverage, i.e. to be as tight as possible containing the entire instance of the object. To ensure accurate annotations, the process starts with an initial training that consists of reading a set of instructions, understanding the rules, and passing an assessment test before the participants can engage with the annotation task, as described in detail below.

\subsubsection{Annotation Instructions}
The instructions are composed of the following items. First, we asked students to read a document that outlines the collection of all the 11 intended UI design components (e.g., background image, icon, input field, etc.), together with their functionality and style guide. For example, we included the set of icon concepts extracted in ~\cite{liu2018learning} as an icon style guide to recognize icon patterns and distinguish them from normal images. This enforces a deeper understanding for each aspect of the design elements. Second, we provided an example set of annotated UI design images where all the instances of the UI components already have a bounding box associated with a class label. This provides a better understanding of how to make a good bounding box annotation. Third, we gave specific instructions on how to use the \textit{RectLabel}\footnote{https://rectlabel.com} annotation tool. Students were compensated for the tool's monthly subscription fees. 

\subsubsection{Annotation Rules}
\begin{figure}
  \centering
  \includegraphics[width=1.0\columnwidth]{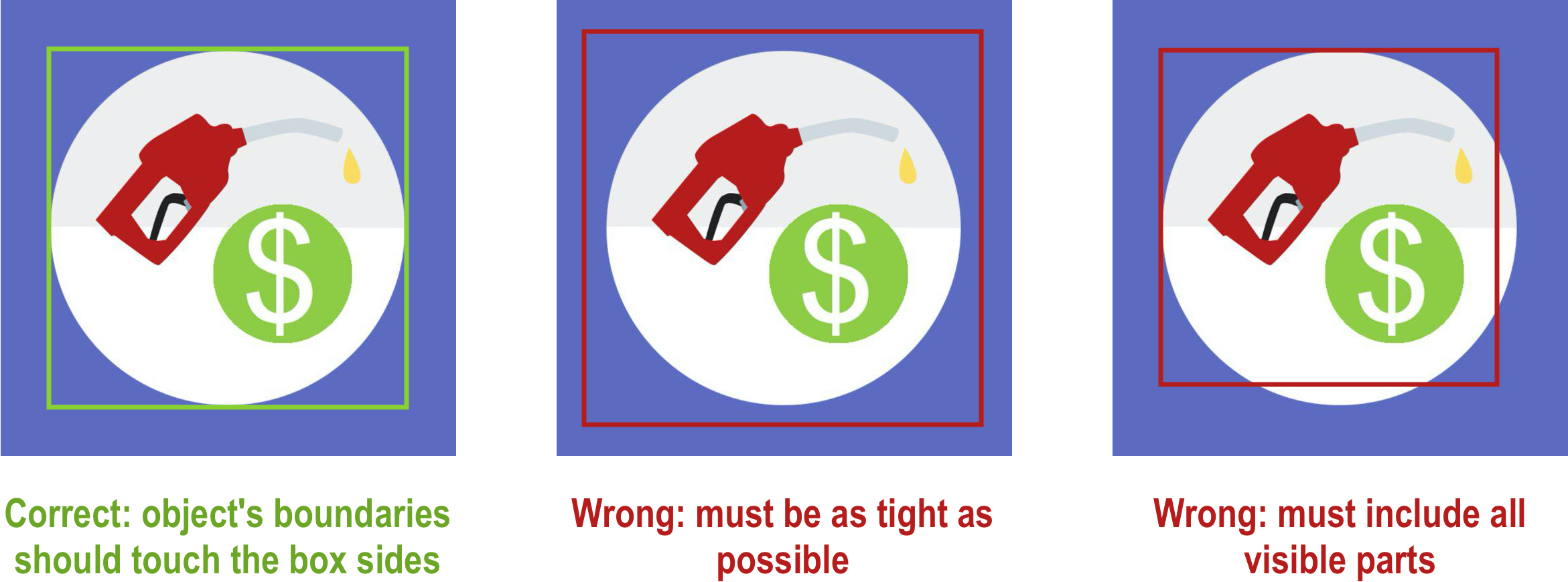}
  \caption{Instructions of drawing a perfect fit bounding box.}~\label{fig:PerfectFit}
  \Description{This figure demonstrates the instructions of drawing a perfect fit bounding box. It contains 3 examples: 1 represent a correct annotation while the other 2 represent wrong annotations. As in the first example, the correct annotation is when the bounding box is as tight as possible perfectly containing the object to be annotated. However, for the second example, it represents a wrong annotation of a loose bounding box which contains much space and unnecessary parts from the background. Finally, the third example represents a wrong annotation of a tight box that does not cover all the visible parts of the object}
\end{figure}

We also provided a set of rules to be followed during the annotation process:
\begin{itemize}
\item \textbf{\textit{Perfect fit}}: When drawing the bounding box, it should be as tight as possible perfectly containing the object to be annotated. It is important to note that the boxes should neither be too tight (i.e., does not cover all the visible parts of the object)  nor too loose (i.e. contains much space and unnecessary parts from the background) as shown in Figure~\ref{fig:PerfectFit}. 

\item \textbf{\textit{Correct Labeling}}: Once the bounding box is drawn, a label must be assigned to it. The label must match the class of the annotated object. To overcome any confusion when assigning the labels, students must refer to the style guide documentation or contact the research team for feedback. 

\item \textbf{\textit{Multiple Objects}}: A bounding box must be drawn for each of the multiple UI objects available in the design image and assigned the correct label accordingly.
\end{itemize}

\subsubsection{Annotation Assessment Test}
We then asked the students to pass an assessment test, which includes a set of test images. Test images were selected to cover all the classes of UI components. Also, we already have the ground-truth bounding boxes for these images and used these to evaluate the quality of the students' annotated results. Within this test, students were requested to complete two tasks: draw new bounding boxes and modify existing ones. For the first task, we provided students with design images that do not yet have bounding boxes and requested them to draw the boxes on the available components. For the second task, we provided students with images that contain bad bounding boxes. These images have been generated by either changing some of the bounding boxes class label or perturbing their coordinates. They were requested to modify these bounding boxes accordingly. 

To ensure quality bounding boxes, students must achieve a 90\% Intersection over Union (IoU) for all images in both tasks. This is an iterative process in which, after each submission, we reviewed the results and provided the students with the necessary feedback if the bounding boxes are not correctly drawn or do not belong to their respective class. They could only start working on the actual images after completing the training with a high IoU score.

\subsubsection{The Annotation Task} 
The process workflow continues with the drawing task, where we gave each student a batch of 100 images and asked them to fully annotate each image. It normally takes an average of 10 hours to complete each batch. Once the drawing task was completed, they proceeded to the task of quality verification. We randomly assigned each student a different batch of fully annotated images from the drawing task and requested them to examine them. They needed to verify the quality and label for all the bounding boxes within an image and modify accordingly. Finally, after completing the quality verification task, we conducted a control verification task on all the annotated images to ensure the quality and coverage of all the bounding boxes. In this task, we needed to examine the bounding boxes quality for each image and modify it accordingly if needed. However, mostly all images were highly annotated and only a few needed slight modifications of the bounding boxes coordinates. 

After the annotation process was completed, the generated VINS dataset contains pairs of a UI design image and its corresponding XML file in the Pascal-VOC format \cite{everingham2010pascal}. To the best of our knowledge, this is the first annotated publicly available UI design dataset for object detection.

\section{Visual Search System}
\begin{figure*}
  \centering
  \includegraphics[width=0.85\linewidth]{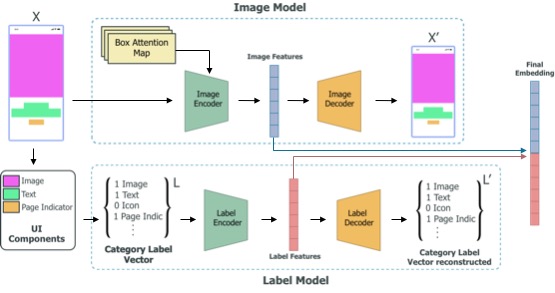}
  \caption{The framework of our image retrieval model. It consists of two parts: an image model and a label model. The image model learns a visual feature vector that encodes the hierarchical structure of the input image. The image encoder is conditioned on a box attention map for better structure learning. The label model learns a feature representing all the available UI components which serve as a high-level control to support the learning process of the visual features. We concatenate the image and label features to produce a final embedding that is used for the retrieval process.}
  \Description{The framework of our image retrieval model which consists two parts: an image model and a label model. The image model is an attention-aware image autoencoder that takes a segmented image as input and learns a visual feature vector that encodes the input’s hierarchical structure. The image encoder is conditioned on a box attention map for better structure learning. The label model is a label encoder-decoder that learns a content feature vector representing all the available UI components which serve as a high-level control to support the learning process of the visual features. The visual and content feature vectors are concatenated to produce a final embedding that is used for the retrieval process. Extra details are included in the text.}
  \label{fig:architecture}
\end{figure*}

VINS, our proposed visual search framework, takes a UI layout as its input query and provides structurally similar UI design examples for inspiration. Instead of just finding similar images that are indexed by their visual content such as color, texture, and shapes, we focused on developing a more advanced visual search system that indexes the image by its functionality (defined by its content) and leverages its structural information. VINS has two main components: \textit{Detection} and \textit{Image Retrieval}. The detection process detects the input query's different UI components to produce a tentative segmented layout. Trained on these generated segmented layouts, our image retrieval process learns a joint feature embedding to find designs similar to the input query. Below is a detailed discussion of the two components.

\subsection{Detection}
To find structurally similar UI design images in the reference dataset, we need to identify and locate the different UI components that exist in the image. Usually, the placement and functionality of these UI components vary widely across the different designs. Therefore, the first step in the process is to accurately infer the bounding boxes of the different UI components and their domain specific types. To achieve this goal, we adopted a Single Shot MultiBox Detector (SSD) model~\cite{liu2016ssd}. We opted to use SSD because of its simplicity and state-of-the-art performance for object detection~\cite{liu2016ssd}. SSD requires taking a single shot to detect the multiple objects within the image, meaning that the tasks of object localization and classification are completed in only a single forward pass of the network. As part of the SSD, we used MobileNets~\cite{howard2017mobilenets} as the base network for feature extraction because they are optimized primarily for speed.

To effectively support the retrieval process of various UI design examples, the object detector has been configured to detect and classify the most common UI components for an input query.  Detector outputs are used to generate a semantic structured layout where each detected bounding box is represented with a unique color, based on its class label. The generation of these semantic layouts provides an easy understanding of the context and hierarchical structure of the input query. This new set of generated semantic layouts represents the collection of images used to train the image retrieval system described below.

\subsection{Image Retrieval}
For each query image, after detecting its different UI components from the previous phase, the next step is to search for design examples in the reference dataset that share a similar hierarchical structure. We believe it is important to focus on two main aspects of each query image: the collection of its UI components and their spatial location. To gain more insight into the image's components, we introduce a high-level attribute encoding the components' class labels. This allows us to learn joint features to measure the semantic relevance across different images at a more fine-grained level.

We propose a multi-modal embedding framework that learns joint features of the image's structure and associated content and uses them to guide the UI retrieval process. The proposed framework consists of two models as shown in Figure~\ref{fig:architecture}. The first model is an attention-aware image autoencoder $E$ that takes an image as input $x$ and produces a structural feature vector $z_{1}= E(x)$ to encode the hierarchical structure of the image. The second model is a label encoder-decoder $A$ that learns a feature vector $z_{2}=A(y)$ capturing the UI component's $y$ in the image. These two feature vectors are fused by concatenation to form the final representation $z = (z_{1},z_{2})$. Below we discuss the two models in detail.

\subsubsection{Image Model} 
\label{section:ImageModel}
Our encoder takes in a semantic input image, which is down-sampled to size 256x256. For a fair comparison of results, we follow the design of the autoencoder described in~\cite{liu2018learning}. It consists of 4 convolutional layers at increasing feature size. The convolutional layers are arranged as 3x8, 8x16, 16x16, 16x32 (input channels x filters). The size of kernel and stride across all the layers is maintained at 3 and 1 respectively. A RELU activation and a max pooling layer of size and stride 2 is applied after every convolutional layer. This is considered the base autoencoder model.

One of our contributions is augmenting the base autoencoder model with a Box Attention mechanism following the approach presented by Kolesnikov et al.~\cite{kolesnikov2019detecting}. This approach was used to model object interactions in an object detection pipeline. We, however, employ it differently by incorporating it in our encoder model to guide the image retrieval process by providing a better comprehensive understanding of the image structure. The idea of this box attention mechanism is to create a map represented as a spatial binary image encoding the location of the UI components. The binary image is of the same size as the original input image with 3 channels. The first channel represents the bounding boxes of the UI components having all the pixels inside the bounding boxes set to 1 and all other pixels are set to 0. The second channel is all zeros, and the third channel is all ones. To integrate this additional box attention into the base encoder model, the attention map is conditioned on the output of the convolutional layers. This conditioning procedure can be applied to every convolutional layer of the base encoder model. However, in our case, we created different models, each employing the above conditioning procedure on a varying number of convolutional layers and reported the performance of each in the results section. The final structural feature vector of the encoder model is represented as 32x16x16 dimensional vector.  

The decoder \textit{D} aims at reconstructing the original image from the latent representation \textit{z\textsubscript{1}} such that \textit{D(z\textsubscript{1}}) is as similar to the input \textit{x} as possible. It consists of the same encoding layers in reverse order, with up-sampling layers instead of max pooling layers. 

To train the autoencoder model, we follow the L2 norm of minimizing the Mean Squared Error (MSE) to formulate the loss function of the model. This loss aims to measure how close the reconstructed input  $\tilde{x}$ \textit{ =D(E(x))} is to the original input \textit{x}:

\begin{equation} \label{eq:1}
L_{rec}(x) = \left \|  x - \tilde{x} \right \|_{2}^{2}\textrm{}
\end{equation}

To further improve the model learning, we introduce a new additional term into our loss function based on the Dice coefficient~\cite{sorensen1948method}. The dice coefficient is an overlap based metric widely used in segmentation problems for pairwise comparison between two binary segmentations. It is based on the intersection over union (IoU) measure and aims at detecting object boundaries to measure the overlap between two samples. it is defined as: 

\begin{equation}
    DiceCoef = \frac{2 * \sum_{i} x_{i} \tilde{x}_{i}}{\sum_{i} x_{i} + \sum_{i} \tilde{x}_{i} }
\end{equation}{}

And the Dice loss function is simply:

\begin{equation}
    L_{Dice} = 1 - DiceCoef
\end{equation}{}

The loss function that our autoencoder aims to minimize is then:
\begin{equation}
L_{AE}(x) = L_{rec} + L_{Dice}
\end{equation}

\subsubsection{Label Model} 
\label{section:LabelModel}
Given an image, we consider the unique class labels of the UI components associated with it. This information is reflective of the content of the image and can serve as a high-level control to support the learning process of the image’s UI layout. In some cases, a specific UI component dominates the image by taking a significant proportion of its layout, which diminishes the rest of the components present. Encoding the class label of the UI components can influence the learning of the overall layout structure. Thus, the label model is built around the encoder-decoder paradigm and it aims to encode the class labels of the detected UI components to convey the UI content. 

Each image is assigned a multi-class label representing the unique classes in the image and is encoded as a multi-hot vector of size 11, where the presence of each class is set to 1. This vector is then fed to a series of 3 fully connected layers of sizes 16, 32, and 64 respectively, which will form a 64-dimensional content vector. As part of our fine-tuning process, we found that a vector of size 64 yields the best results. We used the MSE loss function defined in eq~\ref{eq:1} for training the label encoder. 

Both the image and label models were trained end-to-end until convergence. They were optimized using Stochastic Gradient Descent with a fixed learning rate of 0.00005 and a mini batch of size 32. We selected these hyper-parameters empirically based on the training dataset.

For each query image, the retrieval task focuses on returning a ranked list of the most likely similar images from the reference dataset. This is achieved by estimating the similarity of two images based on the learned embedding vector associated with each image. The embedding vector is a concatenation of both the structural (Section~\ref{section:ImageModel}) and content (Section~\ref{section:LabelModel}) feature vectors. We apply the Euclidean distance measure to estimate the similarity score of the embeddings between the query image and each one of the images from the reference dataset. 
\section{System Evaluation}
We evaluate VINS's performance in a threefold manner.  First, we evaluate the object detection model, then the image retrieval model, and finally the end-to-end combined model.   

\subsection{Object Detection Model}
The first step in VINS is locating and classifying the different UI components in the input query to ensure constructing a good representative layout structure. We trained the employed SSD model~\cite{liu2016ssd} from scratch with a learning rate of 1X10\textsuperscript{-2}. The model performance was evaluated through calculating the mean Average Precision (mAP) and the Area under precision-recall curve (AUC) of all the classes. 

Our first objective is to evaluate the annotations of the VINS dataset against the predefined view hierarchies from the Rico dataset. Since only 2000 images from the Rico dataset have been annotated as part of VINS, we have selected from these annotated images only those containing the most common classes: text, text buttons, icons, images, background images, and page indicators. We made this decision to ensure a fair comparison, as the remaining classes are not sufficiently covered within the images that we selected from Rico. This results in a training dataset containing 1230 images that have been split into a training, validation, and test sets based on 80:10:10 ratio respectively. We select IoU of 0.5 for all the object detection results since its normally considered a good detection ratio.

\begin{table}[]
\begin{center}
\caption{Average Precision at (IoU = 0.5) for the detection of each of the 7 class labels Between Rico's view hierarchies and VINS's annotations.}
\begin{tabular}{|l|c|c|}
\hline
                     & \multicolumn{2}{c|}{\textbf{Average Precision (\%)}} \\ \hline
\textbf{Class Label} & \textbf{Rico Dataset}             & \textbf{VINS Dataset}            \\ \hline
Background Image     & 68.61                     & 89.55                    \\ \hline
Icon                 & 29.61                     & 33.28                    \\ \hline
Image                & 36.13                     & 81.65                    \\ \hline
Text                 & 34.10                     & 71.48                    \\ \hline
Text Button          & 66.91                     & 88.47                    \\ \hline
Page Indicator       & 10.28                     & 63.70                    \\ \hline
Upper Task Bar       & 90.90                     & 90.90                    \\ \hline
\textbf{mAP}         & \textbf{48.08}                     & \textbf{74.15}                   \\ \hline
\end{tabular}
\label{detection-Rico-Ours}
\end{center}
\end{table}

Usually most UI images contain an upper status bar that displays information (e.g. time, battery level, cellular carrier) on the screen's upper edge. As part of our approach, this bar is not being cropped, but rather introduced as a new UI component, for two reasons. First, since it displays information containing text and icons, it is often missclassified as belonging to one of the two classes of text or icons. Second, because our dataset contains different UI styles (e.g. Android, iOS, wireframes) and the position of the upper status bar usually varies depending on the UI, therefore when given a new UI image it can detect the bar's location regardless of its UI style. Although \textit{Rico} dataset contain only Android UIs, we still follow the same convention and include the upper status bar as part of the detection process.

Table~\ref{detection-Rico-Ours} shows the Average Precision (AP) at IoU of 0.5 for each of the 7 classes in both \textit{Rico} and VINS dataset. We can see that VINS's annotations make objects of interest more recognizable to the detector and that we are able to achieve a higher AP across all 7 classes. Overall, VINS's annotations provide more than 26\% increase over the mAP of Rico.

Next, we evaluate the performance of the complete VINS dataset, which contains a total of 4,543 images. We follow the same approach of splitting the dataset into a training, validation, and test sets based on 80:10:10 ratio respectively. We test the model on a test dataset consisting of 450 images and achieve an overall mAP of 76.39\% and AUC of 79.02\% across the predefined set of classes. Table~\ref{detectionT} shows the AP of each of the 12 class labels. Overall, the model has a good performance across most of the classes except with the sliding menu component having the highest AP of 100\%. However, the checked view class has the lowest AP of 44.48\%, which can be attributed to its high cross-class similarity and also size, as object detection models often struggle with detecting small objects~\cite{li2019small}.

\begin{table}[]
\begin{center}
\caption{Average Precision of detection results for each of the 12 class labels on the test set consisting of 450 images from the VINS dataset.}
\begin{tabular}{|l|c|c|}
\hline
                     & \multicolumn{2}{c|}{\textbf{(IoU=0.5)}}                                     \\ \hline
\textbf{Class Label} & \textbf{AP (\%)} & \textbf{AUC (\%)} \\ \hline
Background Image     & 89.33                                & 94.45                                \\ \hline
Checked View         & 44.48                                & 43.70                                \\ \hline
Icon                 & 50.50                                 & 49.55                                \\ \hline
Input Field          & 78.24                                & 80.81                                \\ \hline
Image                & 79.24                                & 81.68                                \\ \hline
Text                 & 63.99                                & 65.30                                \\ \hline
Text Button          & 87.37                                & 92.95                                \\ \hline
Page Indicator       & 59.37                                & 61.07                                \\ \hline
Pop-Up Window        & 93.75                                & 97.20                                \\ \hline
Sliding Menu         & 100                                  & 100                                  \\ \hline
Switch               & 80.00                                & 82.45                                \\ \hline
Upper Task Bar       & 90.40                                & 99.09                                \\ \hline
\end{tabular}
\label{detectionT}
\end{center}
\end{table}

\subsection{Image Retrieval Model}
\subsubsection{Quantitative Evaluation}
To quantitatively evaluate the image retrieval performance, we calculate the precision of the top k recommended images from the ranked retrieved list. To do so, we first remove the wireframe UIs from the dataset since they can’t be retrieved as inspirational design examples resulting in a dataset of 4,543 images. We follow the same aforementioned split approach to create a test set of 450 images.  We recruited 3 interface designers from \textit{Upwork} with considerable mobile UI/UX design experience to assign a label representing the design category for each of the test images. Based on the labels assigned, we identified 8 design groups for the UIs presented as follows: login, login with background image, sign up, introduction, introduction with background image, sliding menu, pop-up window, grid-based, and list-based. We eliminated design groups containing less than 10 related images resulting in a test set of 395 images for this evaluation. Then, we take each image in the test set as a query and retrieve a ranked list of K nearest neighbors. Finally, we calculate the precision score at top K retrieved images (precision@K), which is defined as the percentage of the K retrieved images in the list that belong to the same design category label. Since our test set is relatively small, we set the maximum retrieval limit of K = 10.   

\begin{table}[H]
\begin{center}
\caption{Precision score at top K retrieved images from the validation set for the baseline model and different models of our proposed method with varying number of attention maps \textit{m} applied.}
\resizebox{\columnwidth}{!}{\begin{tabular}{|l| c c c c c c|}
\hline
        & \textbf{Top 1} & \textbf{Top 2} & \textbf{Top 4} & \textbf{Top 6} & \textbf{Top 8} & \textbf{Top 10} \\
        \hline
\textbf{Baseline} & 88.97 & 88.20 & 84.87 & 83.16 & 82.11 & 80.43  \\
\hline
\textbf{Ours (\textit{m}=1)} & 90.76 & 90.12 & 88.20 & 86.45 & 85.03 & 84.05 \\
\hline
\textbf{Ours (\textit{m}=2)} & 91.53 & 91.53 & 89.03 & 87.99 & 87.05 & 85.71 \\
\hline
\textbf{Ours (\textit{m}=3)} & 91.53 & 91.41 & 89.23 & 88.11 & 86.89 & 85.87 \\
\hline
\textbf{Our (\textit{m}=4)} & \textbf{92.05} & \textbf{91.79} & \textbf{89.48} & \textbf{88.37} & \textbf{87.21} & \textbf{86.48} \\
\hline
\end{tabular}}
\label{precT}
\end{center}
\end{table}

\begin{figure*} 
  \centering
  \includegraphics[width=\linewidth]{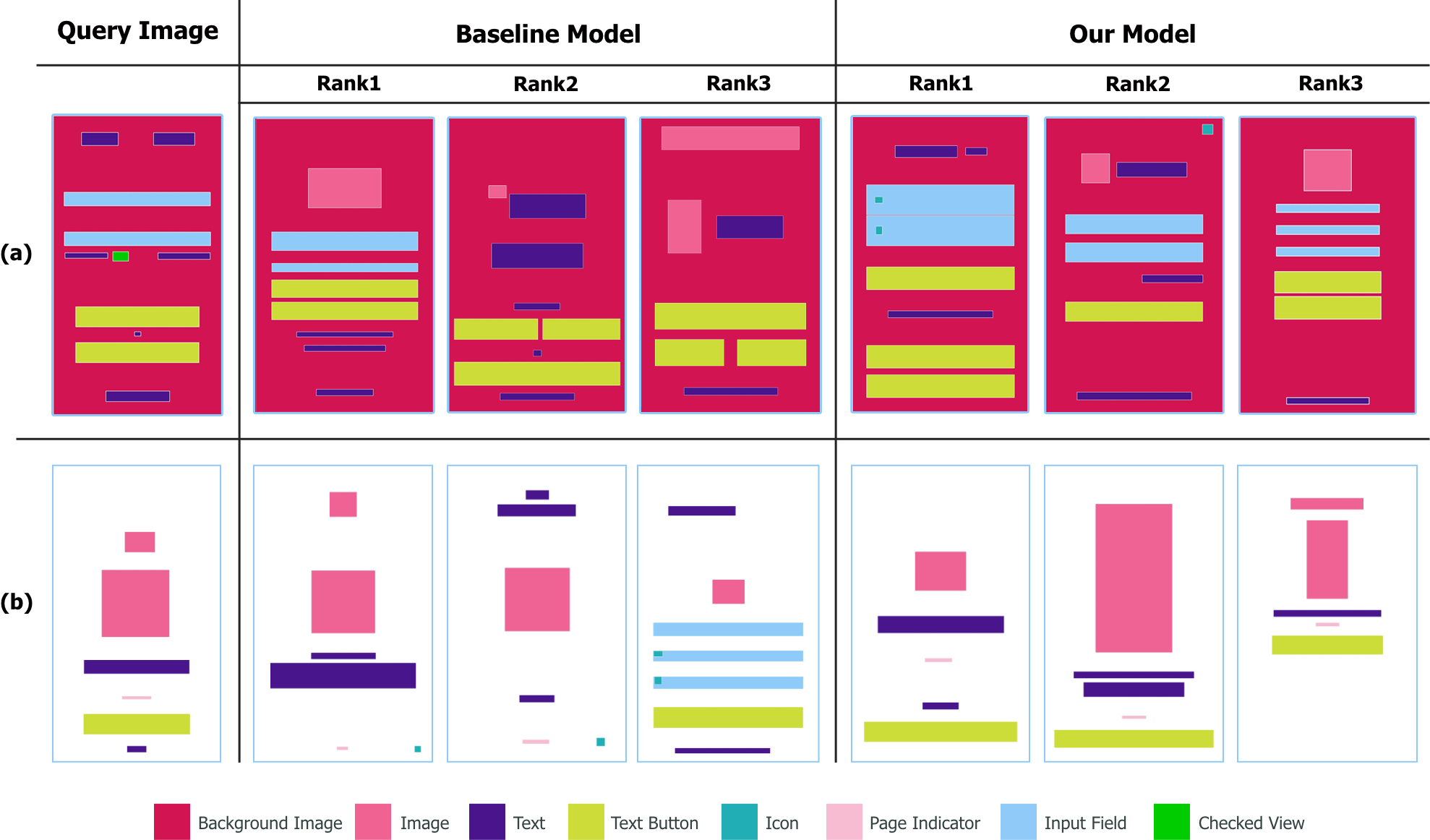}
  \caption{Retrieval results of query images selected from the test dataset from the baseline model and our model. The first column shows the query UI represented as a semantic layout, and the rest are the Top-3 retrieved UIs from both models. The colored bounding boxes represent the different UI components, which are described in the legend below the comparison.}
  \Description{Top-3 retrieval results of 2 query images (example a & b) selected from the test dataset from both the baseline model (First column) and our model (Second Column). In the first row, (example a) is a UI design containing 5 components: background image, text, text button, input field, and checked Text View. In the second row, (example b) is a UI design containing 4 components: image, text, page indicator, and text button arranged in a specific order. Evaluation of the retrieval results from each model is detailed in the text in section 6.2.2 Qualitative Evaluation}
  \label{fig:figure3}
\end{figure*}

In our proposed model, we incorporate the attention box mechanism into our base encoder by conditioning it on the output of the convolutional layers. We treat the attention map \textit{m} as a hyper parameter and experiment by creating different models of varying number of maps. When \textit{m} =1, the model has 1 attention map applied to the output of the last convolution layer. When \textit{m} increases, the number of \textit{m} layers increase and are applied to the last \textit{m} convolutional layers. Because we have 4 convolutional layers, when \textit{m} = 4, each layer has an attention map applied to it.

We compare the retrieval results of our proposed model with the baseline autoencoder implemented in~\cite{liu2018learning}. Table~\ref{precT} shows the precision scores of different number of k neighbors on the validation set for the models used in the experiment. We can see that our model outperforms the baseline and further improves the precision rates at each value of K. The model with \textit{m} = 4 is the best in the overall performance. It is able to achieve 92.05\% precision for the top 1 nearest neighbor and 86.48\% for the top 10 nearest neighbors. This marks an almost 4-6\% improvements over the baseline model. This demonstrates how incorporating these attention maps aids in capturing the UI layout and thus retrieves more relevant images for the input query. We selected the model with \textit{m}=4 and used it to complete the remaining analysis. 

\subsubsection{Qualitative Evaluation}
To qualitatively evaluate the retrieval process, we visualize query results of randomly sampled images from the test set. Although both models often perform similarly, there are cases where our model outperform that baseline in providing examples that better fit the query's content and structure as shown in Figure~\ref{fig:figure3}. 

\begin{figure*} 
  \centering
  \includegraphics[width=\linewidth]{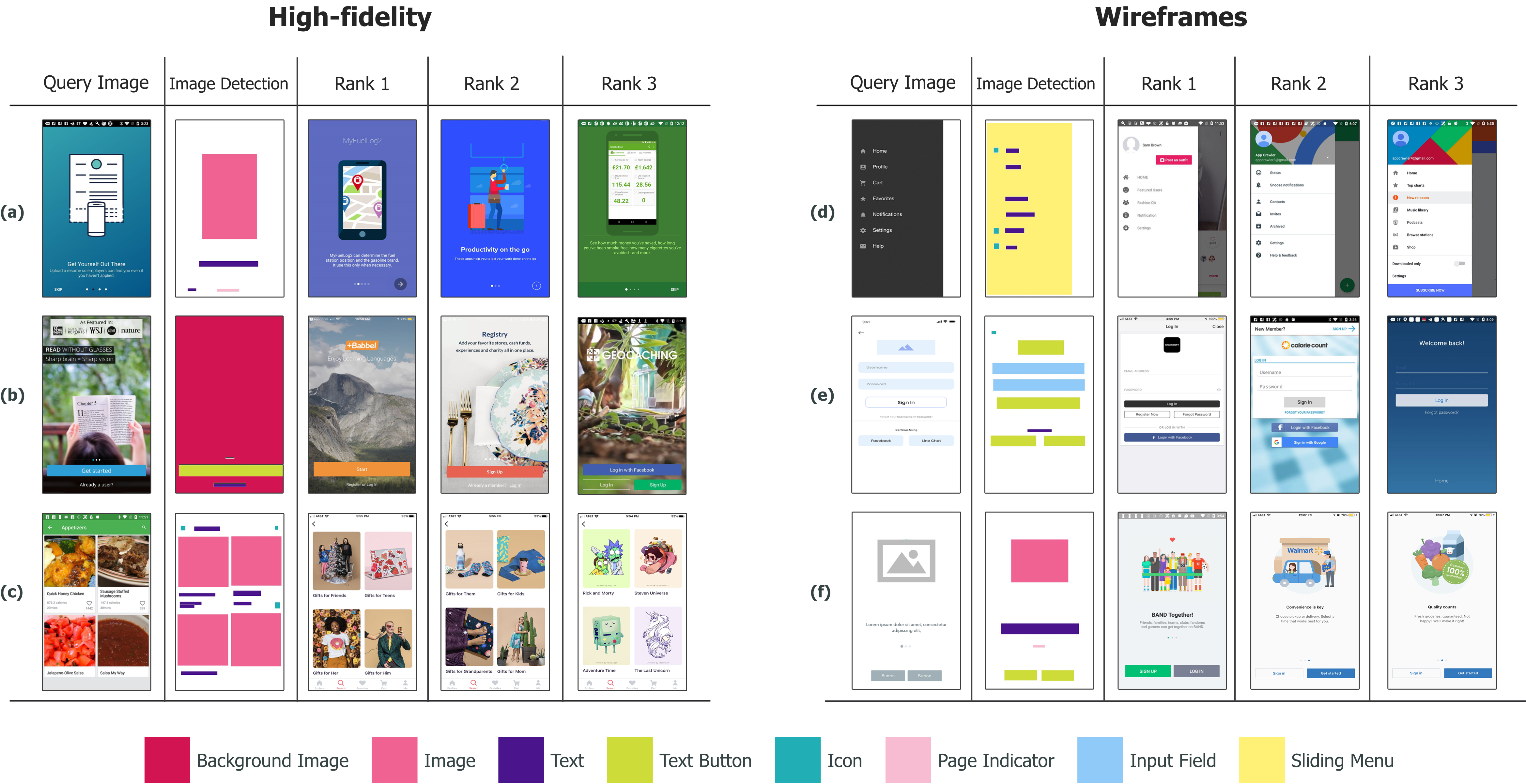}
  \caption{Query Results for VINS, which takes an input query that can be of different design stages (abstract wirframes or high-fidelity designs) (first column), detects its UI components to generate a segmented layout (second column) and then retrieves a ranked list of similar designs (remaining columns).}
  \Description{The Top-3 retrieval results from VINS taking input queries of different design stages showing High-fidelity examples on the left and wireframe examples on the right. It contains 6 examples as follows: (example a) is a high-fidelity onboarding UI with a central image with below text and page indicator, (example b) is a high-fidelity UI page with a background image and login button options at the bottom. (example c) is a high-fidelity UI showing images and their corresponding information in a grid-layout format, (example d) is a wireframe sliding menu UI, (example e) is a wireframe login screen with input fields and login buttons, and (example f) is a wireframe onboarding UI with a central image and two buttons at the bottom. For each example, the figure shows the detection results of its UI components and the Top-3 retrieved layouts which are similar to the input query. Evaluation of the retrieval results is provided in text.}
  \label{fig:CompleteResults}
\end{figure*} 

As part of the evaluation, we consider that the UI components given in the query image are important to the designer. This is derived from the responses of the designers as part of the interview in Section ~\ref{section:interview}. When asked if given a query of a specific layout with a specified number of basic components, 18 designers indicated that a page with exact basic components with various/additional components would be considered an acceptable UI design example.

Our model is able consider the query’s UI components and retrieve examples that fits the overall layout structure (Example a). The baseline, however, retrieves only Rank 1 similar to the query and fails in retrieving the other two because it disregards the input fields and only detects the background image and the position of the text buttons. This may indicate that the baseline model sometimes struggles in understanding the available UI components and may be performing the retrieval based on the dominant color available. We note that both models fail to capture the checked view component from the query, which may have happened because we are validating with a relatively small dataset and, as a result, this dataset may not contain similar designs that have all the exact same components. 

Our model is also able to capture the representation of the query's layout structure (Example b). All ranked results exhibit a consistent structure of following the sequence of component placement in terms of image, text, page indicator, and text buttons, respectively. However, this is not the case for the ranked results from the baseline model as they are either missing a text button for rank 1 and 2 or missing a page indicator and introducing 4 input fields for rank 3. These examples show how our model is better able to identify the UI components and confirm to the query's overall layout structure. 

\subsection{End-to-End Combined Model}
To qualitatively analyze VINS's performance, we visualize the end-to-end query results, including detection and retrieval, on the test set. We discuss below the feedback of expert designers and the potential usage of VINS in supporting other design applications. 
 
\subsubsection{Expert Evaluation}
To gain insight from a professional user's perspective regarding VINS's performance, we recruited 5 new designers from \textit{Upwork}\footnote{https://www.upwork.com/}, to evaluate selected results from the test set. The participants were all US-based UI/UX designers with an average of 2 years of experience and having an average of 85\% job success rate. As part of the evaluation study, we provided them with 10 sets of query designs each containing the corresponding Top-5 retrieved results from the test set. We sampled 5 high-fidelity layouts and 5 abstract wireframes as part of the set to ensure a diverse collection across the design stages. Designers were compensated with \$20 USD for the evaluation study, which lasted approximately 50 minutes.

We asked designers to answer a survey consisting of 5 free-form questions regarding overall relevance between the query and each of the Top-5 retrieved results, the layout and functionality relevance, the effect of additional components in the retrieved set, and whether the design examples provide useful design variations. Some of the specific questions included: ``How would you comment on the relevance between the query and each of the 5 images in the retrieved results?'', ``How would you comment on the layout and functionality relevance between the query and the set of retrieved results?'', and ``Does the set of retrieved results provide useful design variations?''. Similar to the formative interviews (see Section~\ref{section:interview}), two researchers engaged in analyzing this data, with a third researcher confirming the results. 

All the designers mentioned that all the retrieved results are relevant to the input queries and would provide beneficial design examples. As E1 stated “They are very relevant to what was intended in the query”. Specifically, 1 designer appreciated how the layout composition and design patterns of the retrieved results matches the query. As E3 mentioned “I think results looks great based on image query. The layout composition and design patterns elements matches query”. Another designer (E1) mentioned how VINS is able to retrieve similar layouts to the query in example f “Rank 1 to Rank 5 are quite similar to what was required in the query with an image on top and one or two buttons according to the requirement”.  Although the retrieved layouts may be very similar to the query, they do still provide design inspirations such as color schemes as E2 said “Mostly the layout is the same, but they do offer different designs using different color patterns”. In addition to relevancy, all designers agreed that VINS also provides useful design variations regarding different aspects such as “I see properly [sic] designs and all provide useful design variation” (E2), “Yes the design layouts are quite useful” (E1), and “I think this query provides great variations of composition layout” (E5). 

Figure~\ref{fig:CompleteResults} shows 6 different query UIs, which were part of the given survey, and their corresponding Top-3 retrieved results. All designers were satisfied with the design examples for the onboarding queries in regard to design variations in example a: “I really like the variations of the provided examples. Colors, layouts, typography looks great” (E3); and in example f: “they do offer different designs using different color patterns” (E2). One designer (E1) identified the addition of new components such as the forward button in example f. Although there are slight variations in the sliding-menu results (example d), E4 commented that these variations are “beneficial to generate more ideas on how to solve certain design problem”. Designers also observed VINS's capability of detecting the background image and identified the results as “All are unique due to background images” (E2) and “They all offer login functionality with different layouts” (E1). Furthermore, designers expressed satisfaction for the grid-based layouts (example c), providing matching layouts while offering design variations (E1), and providing extra functionality (E4). For the login screen (example e), 4 designers liked the variety in offering different login layout designs and login options. However, E1 commented that some results offer fewer components than what is required in the query, such as that the concept of the google login button is ignored. 

In general, we observed that including extra components provides new ideas and design variations that are appreciated by the designers. As stated by E3 “Adding additional components in query results is definitely beneficial for inspiration and providing possibly better solution”. Similarly, E5 mentioned “They are useful. We might get a new idea after watching extra element”. However, in some cases these extra components may or not be required depending on the condition and the client. We also attempted to understand designers' perspectives regarding the slight variations in either layout or functionality in the retrieved results. E1 identified that layout variations are helpful “For the most part layout matches the query however the functionality is not so. It is not necessarily bad thing because seeing different examples may generate more ideas on how to solve design problem”. E5 however emphasized the importance of functionality over layouts “I think layout doesn't matter here. Our priority should be given the right functionality”. The rest of the designers identified that the layout and functionality of the results match the query. This reflects VINS's effectiveness in retrieving useful design variations and how it is important within the searching tool to have a balance between these two design aspects: offering similar layouts while maintaining the functionality.

\subsubsection{Auto Completion of UI Layouts}
\begin{figure}
  \centering
  \includegraphics[width=\columnwidth,height=\textheight,keepaspectratio]{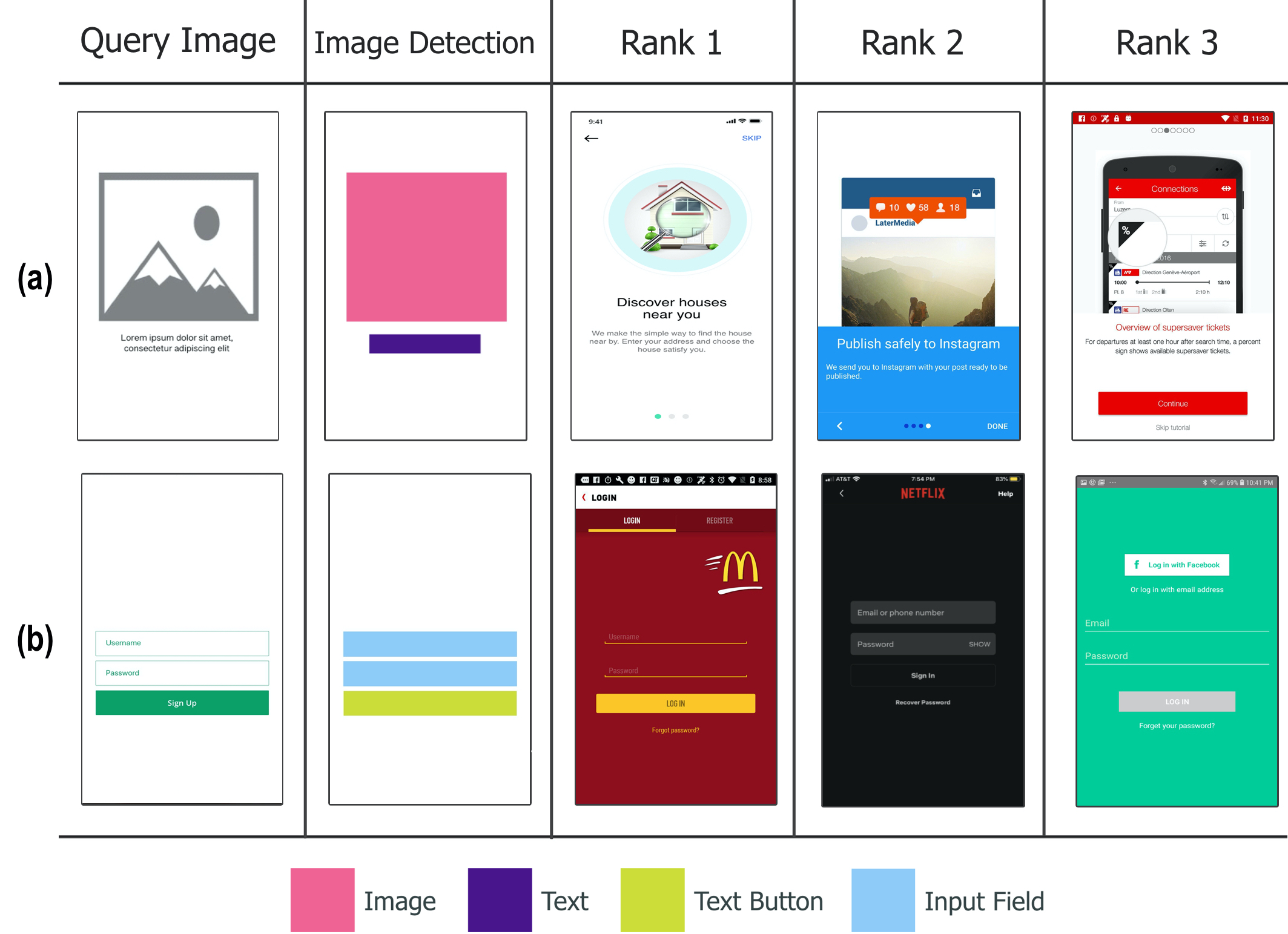}
  \caption{Auto completion of UI layout design. The framework is able to provide design examples that complete the remaining UI components based on partial layouts provided by the user.}
  \Description{VINS’s Retrieval Results of an input query that is a partial UI layout containing only 2-3 UI components. Contains two examples: (example a) a UI design containing a central image and text placed on top of each other, and (example b) a UI design containing two input fields and a text button. For each example, the figure shows the detection results of its UI components and the Top-3 retrieved layouts which are similar to the input query.}
  \label{fig:PartialUI}
\end{figure}

Although VINS sometimes fails to detect all of the UI components, we observed its ability to retrieve examples similar to the partial detected layout. This is related to the problem of auto completing UI layouts~\cite{li2016auto}. To evaluate VINS in supporting this task, we created partial abstract wireframe layouts containing only 2-3 UI components. Based on the partial layout given, VINS brings design knowledge to the process by identifying the entered UI components and then suggesting design examples that complete the layout. As shown in Figure~\ref{fig:PartialUI}, VINS provides design examples that maintain the common components of the query (i.e., central image and text) while providing inspiration on how to complete the remaining UI components (example a). VINS also provides ideas on how to complete a certain layout, e.g. login, by retrieving various UIs with the detected input fields and text button (example b).

These examples illustrate the potential of VINS to facilitate the design process by assisting designers to create layouts. While \textit{Swire}~\cite{huang2019swire} also supports the auto-completion of partial designs, it cannot be implemented directly. Their approach relies on training an alternative model on a new training set with only partial sketches, which is computationally expensive and impractical. VINS can, however, directly accept partial layouts and provide design examples accordingly.

\section{Discussion}
While VINS can retrieve highly similar relevant UI layouts, and marks a clear improvement over previous approaches~\cite{liu2018learning}, there are several limitations, improvements, and future directions that we discuss below.

\subsection{Increasing the Dataset}
Utilizing deep learning frameworks requires a dataset. As existing datasets were unavailable or not sufficient for the purpose of our work, we proceeded in collecting and annotating the VINS dataset, a large mobile UI dataset consisting of UI screens across different design stages. While the VINS dataset was large enough for this work, we still observed that it was not always able to detect all the UI components or provide similar designs. An even larger dataset could potentially address this through improving the detection performance and providing more design variations. The VINS dataset is already publicly available, and we will provide suggestions on how others can include new UI screens so that the dataset can be increased over time. 

Furthermore, it currently only includes 11 classes of the most common UI components for the detection process, which limits the applicability of our model to certain UIs with a set of defined components. This can be improved by including additional UI components spanning different functionalities and identifying the different input field labels (e.g., password, email, etc.), text button concepts (e.g. login, skip, etc.), and icon classes (e.g. social media, settings, etc.) as identified in ~\cite{liu2018learning}. Understanding the true nature of components will aid in generating a more fine-grained hierarchical structure that better defines the UI layout and its design components and thus retrieve even more similar results. This dataset can also be utilized for other data-driven applications such as mobile layout generation~\cite{li2019layoutgan} and UI code generation ~\cite{nguyen2015reverse, moran2018machine, beltramelli2018pix2code}. 

\subsection{Improving Visual Search}
VINS consists of two main components: object detection and image retrieval, both of which can be improved. Although the object detection was able to detect certain classes with very high precision, it failed to do the same for other classes such as checked view and icon. This is due to the high cross-class similarity and large in-class variance. It is also related to the issue of missed detection of the SSD model in small object detection, which can be improved by including more images, using augmentation techniques, or utilizing feature pyramid network structure to enhance detection~\cite{li2019small}. 

As indicated by designers within the interview, they focus on three aspects of design: functionality, structure, and visuality. We can improve our image retrieval model to incorporate, along with the structure and content, the visual features of the UI, such as imagery, font~\cite{o2014exploratory}, and colors~\cite{jahanian2017colors, lin2013probabilistic, o2011color}. We can also improve the structural representation of the UI by utilizing a tree-based data structure to encode the hierarchical view of the layout. This ensures a better modeling for the relations between the different UI components. In addition to image retrieval, such tree structures can also be used to automate different design tasks, including auto-completion of partial designs~\cite{li2016auto} and generation of new layouts~\cite{li2019layoutgan}, which we leave for future work.

Because layout is an important factor in graphic design in general, our visual search framework can be extended from mobile app’s to other design layouts including magazines, posters, web pages, etc. 

\subsection{Working with UI Designers}
Although this paper shows VINS's capacity in supporting example finding behavior, the work presented so far needs to be further validated by the respective users. To achieve this, we plan to conduct user evaluation studies with UI designers to assess how it can be integrated with their daily workflow and how it meets their requirements. Such user evaluation falls into the emerging topic of human-AI interaction~\cite{amershi2019guidelines}.

Most designers were very optimistic about VINS. Their feedback in the interviews provided additional insights and unexpected opportunities that can help for future work. One aspect is that VINS can be utilized for the early stages of design as stated by D7: “Yes, it can be helpful by creating your thoughts on papers as many times as you want, to know about the outlooks of the design”. We can easily extend our dataset to include sketches, by utilizing \textit{Swire's} dataset~\cite{huang2019swire} for example. Another interesting aspect is that VINS can be leveraged into the design process to create an interactive experience between designers and clients to enhance the idea communication phase. As reported by D11, “Even a client can be asked to make a layout and see what it looks like”. Designers also suggested the need to provide user-specified constraints, along with a query image, that can control the searching process, such as keywords and colors. 
\section{Conclusion}
In this paper, we proposed an object-detection based visual search framework for UI layout designs. To support the development of our framework, we (1) interviewed UI designers to better understand the problems, needs, and requirements for visual search; and (2) collected a large-scale annotated UI dataset consisting of UI screens across different design stages that can be utilized for Object Detection training. Utilizing this dataset, our framework first takes an app's design image, which can be an abstract wireframe or a high-fidelity image and detects its UI components to construct a tentative segmented layout representation. It then trains a multi-modal embedding model with an attention mechanism on these generated semantic layouts to learn a joint feature representation that can retrieve similar UI designs. Our findings show a promising performance from both the detection and the retrieval phases. We achieved a mAP of 76.39\% for the detection of different UI components and a precision between 80-90\% for the retrieval of relative design examples for an input query. 


\bibliographystyle{ACM-Reference-Format}
\balance
\bibliography{references}


\end{document}